# Bibliometric-Enhanced Information Retrieval: 3rd International BIR Workshop

Philipp Mayr*, Ingo Frommholz**, Guillaume Cabanac***

* GESIS – Leibniz Institute for the Social Sciences, Unter Sachsenhausen 6-8, 50667 Cologne, Germany
`philipp.mayr@gesis.org`

** Department of Computer Science and Technology, University of Bedfordshire, Luton, UK
`ingo.frommholz@beds.ac.uk`

*** Department of Computer Science, IRIT UMR 5505 CNRS, University of Toulouse, 118 route de Narbonne, F-31062 Toulouse cedex 9, France
`guillaume.cabanac@univ-tlse3.fr`

**Abstract.** The BIR workshop brings together experts in Bibliometrics and Information Retrieval. While sometimes perceived as rather loosely related, these research areas share various interests and face similar challenges. Our motivation as organizers of the BIR workshop stemmed from a twofold observation. First, both communities only partly overlap, albeit sharing various interests. Second, it will be profitable for both sides to tackle some of the emerging problems that scholars face today when they have to identify relevant and high quality literature in the fast growing number of electronic publications available worldwide. Bibliometric techniques are not yet used widely to enhance retrieval processes in digital libraries, although they offer value-added effects for users. Information professionals working in libraries and archives, however, are increasingly confronted with applying bibliometric techniques in their services. The first BIR workshop in 2014 set the research agenda by introducing each group to the other, illustrating state-of-the-art methods, reporting on current research problems, and brainstorming about common interests. The second workshop in 2015 further elaborated these themes. This third BIR workshop aims to foster a common ground for the incorporation of bibliometric-enhanced services into scholarly search engine interfaces. In particular we will address specific communities, as well as studies on large, cross-domain collections like Mendeley and ResearchGate. This third BIR workshop addresses explicitly both scholarly and industrial researchers.

**Keywords:** Bibliometrics, Scientometrics, Informetrics, Information Retrieval, Digital Libraries

# 1 Introduction

IR and Bibliometrics are two fields in Information Science that have grown apart in recent decades. But today 'big data' scientific document collections (e.g., Mendeley, ResearchGate) bring together aspects of crowdsourcing, recommendations, interactive retrieval, and social networks. There is a growing interest in revisiting IR and bibliometrics to provide cutting-edge solutions that help to satisfy the complex, diverse, and long-term information needs that scientific information seekers have, in particular the challenge of the fast growing number of publications available worldwide in workshops, conferences and journals that have to be made accessible to researchers. This interest was shown in the well-attended recent workshops, such as "Computational Scientometrics" (held at iConference and CIKM 2013), "Combining Bibliometrics and Information Retrieval" (at the ISSI conference 2013) and the previous BIR workshops at ECIR. Exploring and nurturing links between bibliometric techniques and IR is beneficial for both communities (e.g., Abbasi and Frommholz, 2015; Cabanac, 2015, Wolfram, 2015). The workshops also revealed that substantial future work in this direction depends on a rise in ongoing awareness in both communities, manifesting itself in tangible experiments/exploration supported by existing retrieval engines.

It is also of growing importance to combine bibliometrics and information retrieval in real-life applications (see Jack et al., 2014; Hienert et al., 2015). These include monitoring the research front of a given domain and operationalizing services to support researchers in keeping up-to-date in their field by means of recommendation and interactive search, for instance in 'researcher workbenches' like Mendeley / ResearchGate or search engines like Google Scholar that utilize large bibliometric collections. The resulting complex information needs require the exploitation of the full range of bibliometric information available in scientific repositories. To this end, this third edition of the BIR workshop will contribute to identifying and exploring further applications and solutions that will bring together both communities to tackle this emerging challenging task.

The first two bibliometric-enhanced Information Retrieval (BIR) workshops at ECIR 2014[1] and ECIR 2015[2] attracted more than 40 participants (mainly from academia) who engaged in lively discussions and future actions. For the third BIR workshop[3] we build on this experience.

# 2 Goals, Objectives and Outcomes

Our workshop aims to engage the IR community with possible links to bibliometrics. Bibliometric techniques are not yet widely used to enhance retrieval processes in digital libraries, yet they offer value-added effects for users (Mutschke, et al., 2011).

---

[1] http://www.gesis.org/en/events/events-archive/conferences/ecirworkshop2014/
[2] http://www.gesis.org/en/events/events-archive/conferences/ecirworkshop2015/
[3] http://www.gesis.org/en/events/events-archive/conferences/ecirworkshop2016/

To give an example, recent approaches have shown the possibilities of alternative ranking methods based on citation analysis can lead to an enhanced IR.

Our interests include information retrieval, information seeking, science modelling, network analysis, and digital libraries. Our goal is to apply insights from bibliometrics, scientometrics, and informetrics to concrete, practical problems of information retrieval and browsing. More specifically we ask questions such as:

- The tectonics of IR and bibliometrics: convergent, divergent, or transform boundaries?
- How feasible and effective is bibliometric-enhanced IR in accomplishing specific complex search tasks, such as literature reviews and literature-based discovery (Bruza and Weeber, 2008)?
- How can we build scholarly information systems that explicitly use bibliometric measures at the user interface?
- How can models of science be interrelated with scholarly, task-oriented searching?
- How can we combine classical IR (with emphasis on recall and weak associations) with more rigid bibliometric recommendations?
- How can we develop evaluation schemes without being caught in too costly setting up large scale experimentation?
- Identifying suitable testbeds (like iSearch corpus[4])

## 3      Format and Structure of the Workshop

The workshop will start with an inspirational keynote by Marijn Koolen "Bibliometrics in Online Book Discussions: Lessons for Complex Search Tasks" to kick-start thinking and discussion on the workshop topic (for the keynote from 2015 see Cabanac, 2015). This will be followed by paper presentations in a format that we found to be successful at BIR 2014 and 2015: each paper is presented as a 10 minute lightning talk and discussed for 20 minutes in groups among the workshop participants followed by 1-minute pitches from each group on the main issues discussed and lessons learned. The workshop will conclude with a round-robin discussion of how to progress in enhancing IR with bibliometric methods.

## 4      Audience

The audiences (or clients) of IR and bibliometrics partially overlap. Traditional IR serves individual information needs, and is – consequently – embedded in libraries, archives and collections alike. Scientometrics, and with it bibliometric techniques, has a matured serving science policy.

We propose a half-day workshop that should bring together IR and DL researchers with an interest in bibliometric-enhanced approaches. Our interests include information retrieval, information seeking, science modelling, network analysis, and digi-tal libraries. The goal is to apply insights from bibliometrics, scientometrics, and informetrics to concrete, practical problems of information retrieval and browsing.

---

[4] http://www.gesis.org/fileadmin/upload/issi2013/BMIR-workshop-ISSI2013-Larsen.pdf

The workshop is closely related to the BIR workshops at ECIR 2014 and 2015 and strives to feature contributions from core bibliometricians and core IR specialists who already operate at the interface between scientometrics and IR. In this workshop, however, we focus more on real experimentations (including demos) and industrial participation.

## 5 Output

The papers presented at the BIR workshop 2014 and 2015 have been published in the online proceedings http://ceur-ws.org/Vol-1143, http://ceur-ws.org/Vol-1344. We plan to set up online proceedings for BIR 2016 again. Another output of our BIR initiative was prepared after the ISSI 2013 workshop on "Combining Bibliometrics and Information Retrieval" as a special issue in Scientometrics. This special issue has attracted eight high quality papers and will appear in early 2015 (see Mayr and Scharnhorst, 2015). We aim to have a similar dissemination strategy for the proposed workshop, but now oriented towards core-IR. In this way we shall build up a sequence of explorations, visions, results documented in scholarly discourse, and create a sustainable bridge between bibliometrics and IR.

### References

1. Abbasi, M. K., & Frommholz, I. (2015). Cluster-based Polyrepresentation as Science Modelling Approach for Information Retrieval. Scientometrics. doi:10.1007/s11192-014-1478-1
2. Bruza, P., & Weeber, M. (2008). Literature-based Discovery. Information Science and Knowledge Management series, vol. 15. Berlin: Springer.
3. Cabanac, G. (2015). In Praise of Interdisciplinary Research through Scientometrics. In Proc. of the 2nd Workshop on Bibliometric-enhanced Information Retrieval (BIR2015) (pp. 5–13). Vienna, Austria: CEUR-WS.org.
4. Hienert, D., Sawitzki, F., & Mayr, P. (2015). Digital Library Research in Action – Supporting Information Retrieval in Sowiport. D-Lib Magazine, 21(3/4). doi:10.1045/march2015-hienert
5. Jack, K., López-García, P., Hristakeva, M., & Kern, R. (2014). {{citation needed}}: Filling in Wikipedia's Citation Shaped Holes. In Bibliometric-enhanced Information Retrieval, ECIR. Amsterdam. Retrieved from http://ceur-ws.org/Vol-1143/paper6.pdf
6. Mayr, P., & Scharnhorst, A. (2015). Scientometrics and Information Retrieval - weak-links revitalized. Scientometrics. doi:10.1007/s11192-014-1484-3
7. Mutschke, P., Mayr, P., Schaer, P., & Sure, Y. (2011). Science models as value-added services for scholarly information systems. Scientometrics, 89(1), 349–364. doi:10.1007/s11192-011-0430-x
8. Wolfram, D. (2015). The Symbiotic Relationship Between Information Retrieval and Informetrics. Scientometrics. doi:10.1007/s11192-014-1479-0